\documentclass[12pt]{article}
\input epsf.tex 
\def\beq{\begin{equation}} 
\def\eeq{\end{equation}} 
\def\bea{\begin{eqnarray}} 
\def\eea{\end{eqnarray}} 
\def\bq{\begin{quote}} 
\def\eq{\end{quote}} 
\def\lessim{\mathrel{\mathpalette\vereq<}} 
 
\newcommand{\gsim}{\stackrel{>}{_\sim}}

\newcommand{\drawsquare}[2]{\hbox{%
\rule{#2pt}{#1pt}\hskip-#2pt
\rule{#1pt}{#2pt}\hskip-#1pt
\rule[#1pt]{#1pt}{#2pt}}\rule[#1pt]{#2pt}{#2pt}\hskip-#2pt
\rule{#2pt}{#1pt}}
\newcommand{\fund}{\raisebox{-.5pt}{\drawsquare{6.5}{0.4}}}

\def\CL{{\cal L}}

\def\tbar{\overline \theta} 
\def\Imtr#1{I\hskip-1pt m\, tr\hskip0pt\left[#1\right]}
\def\plum{\mbox{$\mathcal{E} \hskip-.05in\cdot \hskip.05in$}}
\makeatletter 
\def\vereq#1#2{\lower3pt\vbox{\baselineskip1.5pt \lineskip1.5pt 
\ialign{$\m@th#1\hfill##\hfil$\crcr#2\crcr\sim\crcr}}} 
\makeatother 
 
\title{\Huge Strong-weak {\sc CP} hierarchy from non-renormalization theorems%
\footnote{Work supported by the Department of Energy,  
Contracts DE-AC03-76SF00515 (GH) and DE-AC02-76CH03000 (MS)}} 
\author{ 
        Gudrun Hiller $^{a}$ 
        and Martin Schmaltz $^b$\ \\ \\ \\ 
        \small \sl $^a$\ SLAC, Stanford University, Stanford, CA 94309\\
        \small \tt ghiller@slac.stanford.edu \\
        \small \sl $^b$\ Department of Physics, Boston University, 
Boston, MA 02215\\ 
        \small \tt schmaltz@bu.edu \\
\\        } 
 
\begin{document} 
\baselineskip=17pt 
\pagestyle{plain} 
 
\begin{titlepage} 

\maketitle 
\begin{picture}(0,0)(0,0) 
\put(200,350){FERMILAB-Pub-01/379-T} 
\put(253,330){SLAC-PUB-9080} 
\put(267,310){BUHEP-01-35} 
\end{picture} 
 
 \begin{abstract} 

\leftskip-.6in 
\rightskip-.6in 
\vskip.2in
 
We point out that the hierarchy between the measured values of the CKM
phase and
the strong CP phase has a natural origin in supersymmetry with
spontaneous CP violation and low energy supersymmetry breaking.
The underlying reason is simple and elegant: in supersymmetry the
strong CP phase is protected by an exact non-renormalization theorem
while the CKM phase is not. We present explicit examples of
models which exploit this fact and discuss corrections to the
non-renormalization theorem in the presence of supersymmetry breaking.
This framework for solving the strong CP problem has generic predictions
for the superpartner spectrum, for CP and flavor violation, and predicts a
preferred range of values for electric dipole moments.

\end{abstract} 
\thispagestyle{empty} 
\setcounter{page}{0} 
\end{titlepage}

\section{Introduction} 

Despite it's impressive phenomenological success the Standard Model
has serious short-comings which should be understood as pointers
towards physics beyond the standard model. One such shortcoming is
the puzzling hierarchy between the CP violating phase in the
CKM matrix and the strong CP phase $\tbar$. This ``strong CP problem''
\cite{reviews}
has recently become more severe as results from the $B$-factories now
clearly favor a unitarity triangle with three large angles
\cite{PDG,recentckmfit}, implying
that the complex phase in the CKM matrix is of order one.
In contrast, the strong CP phase which is the only other CP violating
parameter in the Standard Model has been experimentally bounded to
be tiny, $\tbar \le 10^{-10}$ from measurements of
electric dipole moments of the neutron and $^{199}$Hg
\cite{strongnEDM,strongnEDMexp,strongHgEDMexp}.

In the Standard Model this hierarchy between the two CP violating phases
is puzzling because the phases have a common origin: the Yukawa
couplings of the quarks.  The CKM matrix is the unitary transformation
matrix which takes one from the basis with a diagonal up quark Yukawa matrix
$Y_u$ to the basis with a diagonal down quark Yukawa matrix $Y_d$.
An irremovable large phase in the CKM matrix implies at least one irremovable
large phase in the Yukawa matrices.
This then requires a fine-tuning of the strong CP phase to one part
in $10^{10}$ because $\tbar$ depends on the phases in the Yukawa
matrices via
\beq
\tbar=\theta-{\rm arg det} Y_u - {\rm arg det} Y_d \ .
\label{eq:thetaeq}
\eeq
Here, we denote the physical (re-phase invariant) theta angle
with $\tbar$ to distinguish it from the basis dependent unphysical
``bare'' $\theta$. 

Several resolutions of the puzzle have been proposed. The axion
mechanism
\cite{axion}
promotes $\tbar$ to a field. QCD dynamics gives this
field a potential with a minimum at zero. Experimental searches for
the axion have come up empty-handed, and -- when combined with
constraints from cosmology and astrophysics -- they have reduced the
allowed parameter space to a narrow window \cite{PDG}.
Another proposed solution, a vanishing up quark mass \cite{mu=0}, is on the
verge of being ruled out by using partially quenched chiral
perturbation theory to compare lattice calculations to
experiment \cite{lattice}.

There are also proposals based on specific models which we may
classify as ``high-scale solutions''
\cite{NB}-\cite{glashow}, the most famous of which is the Nelson-Barr
mechanism \cite{NB}. These models use a symmetry
(parity or CP) to enforce $\tbar=0$ at high scales.
But in the Standard Model both P and CP are badly broken,
and it becomes a challenging and cumbersome model building task to
design realistic models which predict $\tbar < 10^{-10}$ also
at low energies after including all renormalization effects.
While some of the models in the literature work,
they lack the appeal of the axion and $m_u=0$ solution which
attempt to solve the strong CP problem with symmetries at low energies
and are therefore relatively robust against changes in the high-energy
theory and renormalization.

Recently, we pointed out \cite{hilschshort} that by marrying
spontaneous CP violation with supersymmetry one can
construct viable high-scale solutions in which $\tbar$ is automatically
insensitive to radiative corrections and new high energy physics. 
Our proposal makes use of the fact that
in supersymmetry the strong CP phase $\tbar$ is not renormalized
because of a non-renormalization theorem \cite{EFN}.
This makes the task of building a successful model much easier. One
only needs to make sure that $\tbar$ is zero at the tree level.
Loop corrections are automatically absent if supersymmetry breaking
occurs at energies much below the spontaneous CP violation.

In our previous publication \cite{hilschshort}
we briefly introduced our framework and
presented an example model. 
The basic ingredients of the framework are 

\noindent \quad
$\bullet$ \quad spontaneous CP violation

\noindent \quad
$\bullet$ \quad SUSY non-renormalization theorems

\noindent \quad
$\bullet$ \quad flavor and CP preserving
SUSY breaking such as gauge mediation

In this paper we discuss our mechanism
in more detail and provide a number of arguments and calculations
to corroborate the claims made in \cite{hilschshort}.
In particular, in Section \ref{sec:frame} 
we review our general framework. In Section \ref{sec:wave}
we show that a sufficiently large CKM phase can be generated
from wave function renormalization. We also
review the supersymmetric non-renormalization theorem for $\tbar$.
Section \ref{sec:cpx} 
is devoted to explicit models, we discuss a Nelson-Barr
model in which the CKM phase is generated at the tree level. 
We further present a model in which the CKM phase vanishes at tree level
but is generated at the loop level from strongly coupled CP
violating dynamics. 
In Section \ref{sec:susybreak} 
we discuss the spectrum of supersymmetry breaking
masses which is required for a successful implementation of our
scheme. In Section \ref{sec:gmsb} we determine the expected size of
$\tbar$ from radiative corrections in the Standard Model and
from supersymmetry breaking.
Sections \ref{sec:predictions} and 
\ref{sec:summ} contain our predictions, summary and conclusions.
In Appendices A -- D we define our notation, show that a large
CKM phase from wave function renormalization requires strong coupling and
present calculational details regarding the renormalization and
non-renormalization of $\tbar$.

\section{The framework \label{sec:frame}}

In this section we summarize the basic ingredients of our
framework. More details on each will be given in the following
sections.

We require CP and supersymmetry to be exact at high energies. 
At such energies our theory is therefore
described by a supersymmetric Lagrangian with coupling constants
which can be chosen real. We will think of this Lagrangian
as an effective Lagrangian valid up to a cut-off scale which
we call $M_{Pl}$ for convenience. But this scale
could be any other high scale of new physics such as the
GUT scale or the string scale. Our Lagrangian also contains
higher dimensional operators suppressed by the cut-off. Such
operators are also required to be supersymmetric and CP preserving.

\begin{figure}[htb]
\vskip 0.0truein
\centerline{
\epsfxsize=4.3in{\epsfbox{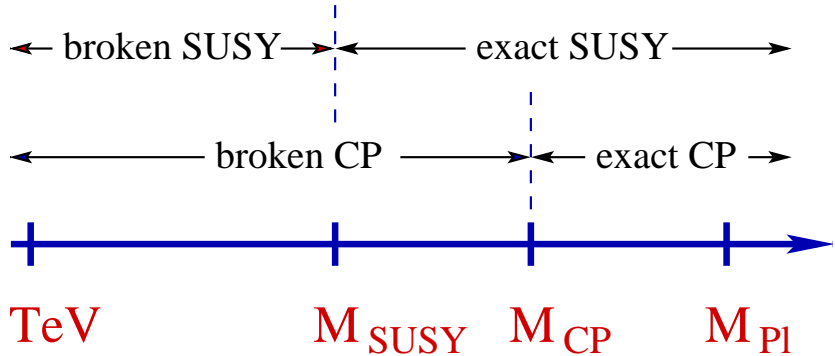}}}
\vskip 0.0truein
\caption[]{\it SUSY and CP breaking scales in our framework.
Figure not to scale. }
\label{fig:pfeil}
\end{figure}

Since the Standard Model is neither SUSY nor CP symmetric
both symmetries must be spontaneously broken.
We denote the scales at which the
symmetry breaking is mediated to the MSSM fields by $M_{CP}$
and $M_{SUSY}$, respectively. Note that this is a somewhat unconventional
definition for $M_{SUSY}$. To be completely clear, in gauge mediation
superpartner masses are proportional to $F/M_{SUSY}$ in our notation,
and in minimal supergravity we would have $M_{SUSY}= M_{Pl}$.

In order for our mechanism to work
we require that $M_{CP}\gg M_{SUSY}$ as shown in Figure
\ref{fig:pfeil}. Therefore the
theory is still supersymmetric at $M_{CP}$ and the well-known
non-renormalization theorems apply. In particular the
strong CP phase $\tbar$ is not renormalized. This makes building
models of spontaneous CP violation which solve the strong CP
problem relatively easy. We only need to require vanishing of
$\tbar=0$ at the tree level, the non-renormalization theorem
guarantees that this remains true after quantum corrections.
However, the CKM phase is renormalized so that a non-vanishing
$\Phi_{CKM}$ can be obtained from quantum corrections
as in our example model of Ref. \cite{hilschshort}
or already at the tree level as in the models of Nelson and Barr. 

At the much lower scale $M_{SUSY}$ Kaehler potential couplings of MSSM
fields to the supersymmetry breaking sector are generated. These
couplings turn into soft supersymmetry breaking masses once
the SUSY breaking fields are replaced by their vacuum expectation
values. It is important that these couplings to
the SUSY breaking sector
do not yet exist at the scale $M_{CP}$. This is because they
would be renormalized and would pick up phases from the CP violating
dynamics. We discuss this issue in more detail in Section \ref{sec:susybreak}.

At scales below $M_{SUSY}$ the theory is simply the MSSM with
soft masses. Thus the low-energy CP violating parameters
can be determined using the well-known renormalization group
equations of the MSSM. This renormalization
only generates negligibly small contributions to $\tbar$ if the soft
SUSY breaking parameters are real and flavor universal. We
review the arguments which prove this in gauge mediated
supersymmetry breaking in Section \ref{sec:gmsb}.

\section{CKM phase from wave functions \label{sec:wave}}

In this section we show explicitly that wave function renormalization
does not contribute to $\tbar$. This is crucial to our mechanism
because wave function renormalization is not constrained by
$N=1$ supersymmetry.
However, and this is important for our model of CP violation
from wave functions in Section \ref{sec:cpx}, wave function renormalization
of the quarks contributes to the CKM phase.
We show that a large CKM phase can be generated entirely
from renormalization of the quark kinetic terms if 
the $Z$-matrices appearing in the renormalization deviate from
the unit matrix by order one. Finally we discuss 
the (non)renormalization of $\tbar$ and $\Phi_{CKM}$
in supersymmetry. 

To begin, consider the following Lagrangian containing the
kinetic terms of the SM quarks and their Yukawa couplings
 
\begin{eqnarray} 
{\cal{L}}_{kinetic}&=&\bar{Q}i\! \not\!\!{D} Z_Q Q+\bar{D} i\! \not\!\!{D} Z_d D 
+\bar{U}i\! \not\!\!{D} Z_u U \label{kinetics} \\ 
-{\cal{L}}_{yukawa}&=&\bar{Q} \hat Y_u H_u U+\bar{Q} \hat Y_d H_d D \ .
\label{yukawas}
\end{eqnarray} 
We use two-component spinor notation, $Q$ are the $SU(2)$-doublet
quarks, $D$ and $U$ are $SU(2)$-singlets. $Z_i$ denote wave function
renormalization factors which in general are complex, Hermitian
and positive definite $3\times 3$ matrices.
Such matrices can always be written as the
square of other positive definite Hermitian matrices $Z_i=(T_i)^{-2}$.
Thus we can always change from this most general basis to canonical
fields by a Hermitian basis change
$Q \to T_Q Q $,
  $U \to T_u U$ and   $D \to T_d D$ 
which leads to new Yukawa matrices 
\begin{eqnarray} 
Y_{u}=T_Q \hat Y_{u} T_{u} ~,\qquad 
Y_{d}=T_Q  \hat Y_{d} T_{d} 
\label{eq:Ychange}
\end{eqnarray} 
It is important to note that this basis change does not shift $\theta$.
This is most easily seen by writing $T=U^\dagger S U$ with unitary
$U$ and real-diagonal $S$. Rescaling
the quark fields by the real matrix $S$ does not change $\theta$
and potential contributions from $U$ and $U^\dagger$ cancel.

It is now easy to see that the contribution to $\tbar$
from quark masses (see Eq.~(\ref{eq:thetaeq})) vanishes
if the only phases in the quark sector are in the $Z_i$.
This follows because
\beq
{\rm arg det} Y_{u/d} =  {\rm arg det} T_Q + {\rm arg det} \hat Y_{u/d}
                    + {\rm arg det} T_{u/d} = 0
\eeq
by hermiticity of $T$ and reality of $\hat Y_{u/d}$.

Note that the phases contained in the $Z$-factors are physical
and lead to a non-vanishing CKM phase. In fact, arbitrary quark
masses and CKM matrices can be obtained
as can be seen from the example
$\hat Y_u = \hat Y_d = T_Q = 1$, $T_u\propto{\rm diag}(m_u,m_c,m_t)$,
and $T_d\propto V_{CKM}\ {\rm diag}(m_d,m_s,m_b) V_{CKM}^\dagger$.

We remark one more result here which is important for the
model in Section \ref{subsec:ourmodel}. In order to generate an
order one CKM phase from wave function factors 
the $Z$'s cannot be close to the unit matrix. In other words if --
for example -- nontrivial $\delta Z$'s (with $Z=1+\delta Z$) are
generated dynamically from loops, then this dynamics needs to be strongly 
coupled so that $\delta Z \sim O(1)$. 
If the $\delta Z$'s are small, then a large CP violating phase cannot be
generated.
While this is plausible, it turns out to be difficult to prove.
A somewhat pedestrian derivation is given in Appendix \ref{app:ckm}.  

To summarize, what we have discussed above outlines a possible
strategy for solving the strong CP problem: if one can construct
a model with vanishing bare $\theta$, real Yukawa matrices,
but complex Hermitian wave function factors $Z_i$, then $\tbar$
vanishes even for large CP violation in the CKM matrix.%
\footnote{We have not yet shown that this is stable under
radiative corrections. We will deal with this in Section \ref{sec:gmsb} where
we discuss renormalization of $\tbar$.}

However, the above is not yet a solution to the strong CP problem.
In the presence of CP-violating dynamics reality of the Yukawa matrices in
Eq.~(\ref{yukawas}) is not enforced by any symmetries,
and it is in general just as miraculous as a vanishing strong CP phase $\tbar$.
But we will show in the following Section that supersymmetry 
and its nonrenormalization theorems can naturally give complex
phases in the kinetic terms and real Yukawa matrices. 

\subsection{Supersymmetry}
\label{sec:susy}

As we will now explain, the situation improves dramatically
in the presence of supersymmetry. This is essentially because in 
supersymmetry $\tbar$ and the Yukawa matrices $\hat Y$ are holomorphic
quantities which are protected by non-renormalization theorems.
However, the wave function factors $Z_i$ stem from the Kaehler potential and
are renormalized. Thus, if it can be arranged in a model that $\tbar$
remains zero at the tree level, then the nonrenormalization theorem
guarantees this also at the quantum level. The CKM phase is
renormalized and can be generated either at the tree level
or by loops.

More explicitly, a supersymmetric Lagrangian can be written as

\beq 
\CL = \int d^4\theta\, K + \int d^2\theta\ W + W_{gauge}\ , 
\eeq 
where $K$ is the Kaehler potential, $W$ the superpotential,
and $W_{gauge}$ contains the gauge kinetic terms. 
Matter fermion masses are given by second variations of the superpotential 
$\hat M_{ij}\equiv \partial^2 W / \partial \phi_i \partial \phi_j$ 
times wave function renormalization 
factors from the Kaehler potential. The wave function 
renormalization is determined from the kinetic terms 
$Z_{ij} \equiv \partial^2 K / \partial \phi_i \partial \phi_j^*$. 
Since $Z_{ij}$ is positive definite and hermitian it can be written 
as the square of a non-singular hermitian matrix $Z=T^{-2}$. 

Chiral superfields in the original basis are related to fields 
in the canonical basis by $\phi_k\rightarrow T_{ki} \phi_i$, and 
the general expression for properly normalized fermion masses is 
\beq 
M_{ij}=- \frac12\, T_{ik}\, \hat M_{kl}\, T_{jl} \ . 
\eeq 
It follows that the contribution to $\tbar$ from $\arg \det M$
vanishes if the couplings (and vevs) in $W$ are real. This remains
true for arbitrary complex Kaehler potential couplings.

In the MSSM, we have
\beq
\hat M_u = \hat Y_u v_u, \quad \hat M_d = \hat Y_d v_d
\eeq
where $v_u$ and $v_d$ are the vevs of the up- and down-type MSSM Higgs
fields, respectively. 
The quark mass matrices are defined in terms of $\hat M_{u,d}$ with products
of wave function factors $T_{Q,u,d}$ as in the non-supersymmetric
case.
 
We are now ready to discuss the non-renormalization of
$\tbar=\theta-\arg\det M$. We showed above that $\arg \det M$ is not
renormalized. To understand the
renormalization of $\theta$ it is convenient to define the superfield
\beq
\tau=\frac1{g^2}+i {\theta \over 8 \pi^2}
\eeq
and work in a basis in which the gauge-kinetic term is $\int d^2\theta
\, \frac14 \tau W_\alpha W^\alpha$, and where no wave function
renormalization is performed. In this basis
$\tau$ is renormalized at one loop only \cite{nsvz,IS}
\beq
\tau(\mu)=\tau(\mu_0)-{b_0\over 8 \pi^2} log(\mu/\mu_0) \ .
\eeq
Here $\mu$ and $\mu_0$ are real renormalization scales
and $b_0$ is the one-loop $\beta$ function coefficient.
Taking the imaginary part on both sides shows that $\theta$\
is also not renormalized.

So far, we have ignored mass thresholds.
A superfield with a mass $m$ between $\mu$ and $\mu_0$
should be integrated out at the scale $m$. This gives a shift
$\delta \tau = - t_2/8\pi^2 log(m)$, and if $m$ is complex
we have $\theta \rightarrow \theta - t_2 \arg m$.
Here $t_2$ is the Dynkin index in the color representation of
the field which was integrated out ($t_2=1$ for a quark).
This is exactly
what is needed for $\tbar$ to be invariant, because the massive field
should not be included in the $\arg \det M$ term in the definition
of $\tbar$ in the low energy theory.\footnote{Note that we have been
somewhat cavalier with the Dynkin indices in the definition of $\tbar$.
The correct definition contains a factor of $t_2(R_i)$ for each of the
different representations $R_i$ of colored fermions in the theory.}
We discuss the non-perturbative generalization of this
non-renormalization theorem in Appendix \ref{app:nonpert}.

To end this section, we wish to clarify a potential
confusion stemming from the possibility
of redefining the phase of the gluino field via an anomalous
R-symmetry transformation. In the absence of a gluino mass this
appears to allow rotating away the $\theta$-angle. However, in order
for supersymmetry breaking to generate a gluino mass as required for
phenomenology, the R-symmetry has to be broken in the theory.
If this breaking is spontaneous then the theory has an R-axion,
and we have re-discovered the axion solution to the strong CP problem
(with it's associated phenomenological constraints). If the breaking is
explicit $\theta$ cannot be rotated away.

\section{The CP violating sector \label{sec:cpx}}

In this section we discuss the requirements on the sector of the
theory which is responsible for breaking CP. We also give two
examples and emphasize the trouble with models without SUSY.

The job of the CP violating sector is to produce a CKM phase
of order one while avoiding a {\it tree level} contribution to $\tbar$.
Quantum corrections to $\tbar$ are automatically taken care of
by the non-renormalization theorem and low energy SUSY breaking
as discussed in Sections \ref{sec:wave} and \ref{sec:susybreak}.

There are two possible strategies for generating the CKM phase.
The first was proposed long ago by Nelson and by Barr. In their
scenario, the ordinary quarks mix with ultra-heavy vector-like quarks
via complex couplings. As we will review in the next subsection
this mixing generates a CKM phase at the tree level while a clever
choice of Yukawa couplings (or equivalently of field content and
global symmetries) forbids the tree level contribution to $\tbar$.
The other possibility was proposed in our recent publication 
\cite{hilschshort}. In our
scenario, CP violation only couples to the MSSM at the loop level.
This automatically guarantees a vanishing $\tbar$, and the CKM phase
must be generated from loops which renormalize the quark wave functions.
We review this scenario in the second subsection.

Spontaneous CP violation requires some fields to have potentials
which are minimized at complex vacuum expectation values.
For simplicity we will omit the specific potentials. They are not
difficult to construct even though the Lagrangian
is real because of the underlying CP invariance. A simple example
is given by the superpotential 
\beq
W= \Xi \, (\Sigma^2+\plum^2)
\eeq
with singlet chiral superfields $\Xi$
and $\Sigma$, whose scalar potential force
a complex vev for $\Sigma=\pm i \plum$.

\subsection{Nelson-Barr}

In Nelson-Barr models \cite{NB} CP violation is communicated to the quarks
already at the tree level. The non-trivial model-building feat
is to arrange the superpotential such that $\tbar=0$ at the tree
level. A relatively simple choice is to add a vector-like 4th
singlet down quark. The superfields $D_4+\overline D_4$ have
$(SU(3),SU(2))_{U(1)}$ quantum numbers
$(\overline 3,1)_{1/3}+ (3,1)_{-1/3}$ and couple in the
superpotential to the MSSM
fields and three complex vevs $\Sigma_i$ as follows
\beq
W= Q_i \hat M_{ij} D_j\quad i,j=1..4\ .
\eeq
Abusing notation, we defined $Q_4\equiv \overline D_4$,
and the mass matrix $\hat M$ is
\beq
\hat M = \left( \begin{array}{ll}
\hat Y_d H_d  & 0 \\
 r\, \Sigma & \mu  
\end{array} \right) \ .
\eeq
Here $\mu\gg M_{weak}$ contributes to the mass of the vector-like fermions,
and all couplings and $\mu$ are real because of the underlying CP
symmetry.
This form of the Lagrangian can be enforced by additional
global symmetries. A similar mixing could also be introduced in
the up-sector and everything is straightforwardly extended to a GUT.

One can easily verify that $\arg \det \hat M =0$, and therefore
$\tbar=0$ at the tree level as desired. The CKM matrix is
the mismatch of the basis in which
\beq
Y_u {Y_u}^\dagger= \hat Y_u {\hat Y_u}^T \quad  {\rm and} \quad 
Y_d {Y_d}^\dagger=
\hat Y_d \left(1-{a a^\dagger \over |a|^2+\mu^2}\right) {\hat Y_d}^T
\eeq
are diagonal. Here we have defined the three-vector
$a^\dagger=(r_1 \Sigma_1,r_2 \Sigma_2, r_3 \Sigma_3)$ and used the
approximation $\hat Y_d H_d \ll r \Sigma \sim \mu$
to compute the down quark Yukawa matrix.
We see that if the vevs $\Sigma_i$ are complex and $a_i \gsim \mu$ the
down quark matrix has large phases giving an unsuppressed CKM
phase as desired.

As discussed in previous sections, supersymmetry guarantees that
$\tbar$ remains zero at the loop level as well. If we had considered
this model without SUSY as originally proposed by Nelson, then we
would have to worry about loops involving the heavy fermions and
$\Sigma$ fields which contribute to $\tbar$. In order to make
the Nelson-Barr models safe without SUSY one needs to take the couplings
$r_i$ very small ($\sim 10^{-3} - 10^{-4}$) while simultaneously
tuning $r_i \Sigma_i \sim \mu$.

\subsection{CKM phase from loops}
\label{subsec:ourmodel}

In this section we review an example model which was 
presented in our first paper \cite{hilschshort}. In this model
the CKM phase stems from wave function renormalization factors
$Z_i$ of the quarks. The $Z_i$ factors arise from loops
of heavy superfields with complex masses. 

Here we will describe an $SU(5)$ GUT version of the model. In
addition to the usual three generations of $\overline 5_i + 10_i$
matter fields we also require a vector-like
$\overline 5_4+5_4$ (models with one or several $10+\overline{10}$
or both $\overline 5+5$ and $10+\overline{10}$ are of course
also possible). Furthermore, we have the usual Higgs $H_u$ and
$H_d$ and three generations of gauge singlet superfields
$F_i+\overline F_i$. The superpotential
contains the usual MSSM couplings as well as
\beq
W_{C\!P \hskip-.1in \backslash}\ = r_{ij} \overline 5_i F_j 5_4 
              + s \Sigma_{ij} F_i \overline F_j 
              + M 5_4 \overline 5_4 \ .
\label{eq:superpot}
\eeq 
$M, r$ and $s$ are real, the indices $i,j$ run over $1..3$, and
the matrix $\Sigma_{ij}$ is assumed to have complex entries
from spontaneous CP breaking.%
\footnote{A more minimal CP violating sector with
only two $F$'s and no $\overline{F}$ would work as well.
Note also that we have vanishing vevs for $F$ and
$\overline F$, with a non-vanishing and complex $r <\overline{F}>\sim M$ 
this model would essentially be Nelsons'.}

\begin{figure}[htb]
\vskip 0.0truein
\centerline{
\epsfxsize=2.6in{\epsfbox{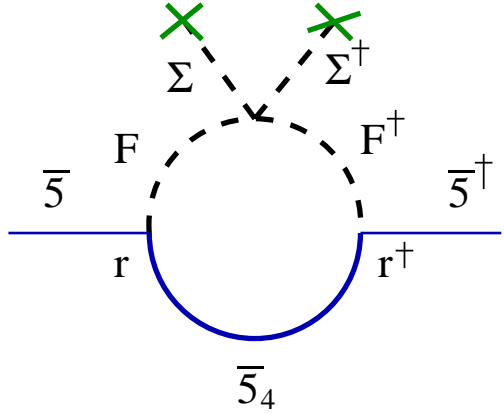}}}
\vskip 0.0truein
\caption[]{\it 
Figure of wave function renormalization for the $\overline 5_i$
following from the superpotential in Eq.~(\ref{eq:superpot}).}
\label{fig:longlegs}
\end{figure}

To determine the low-energy CP violation we integrate out the
massive fields $F$ and $5_4+\overline 5_4$. The
low energy superpotential which derives from Eq.~(\ref{eq:superpot})
vanishes when inserting the equations of motion for $F$'s and $5_4$,
but the diagram of Figure \ref{fig:longlegs} generates
a non-canonical complex kinetic term for $\overline 5_i$. Note that
for CP violation to be mediated to the MSSM fields the
CP breaking sector needs to violate flavor, otherwise the
resulting kinetic term for $\overline 5_i$ would be diagonal and real.
For $\Sigma > M$ the Feynman diagram is easily evaluated.
It's CP violating part involving the vev of $\Sigma$ is finite and
can be expanded to give
 \beq
\delta  Z_{\overline 5} \sim {1\over 16 \pi^2}\ r^\dagger\ 
        {\Sigma^\dagger \Sigma \over M^2_{CP}}\ r 
\eeq 
where $M_{CP}$ is the scale at which spontaneous CP breaking is
mediated to the quarks,
$M_{CP}^2 \sim s^2 tr \left[\Sigma^\dagger \Sigma\right]$.

As discussed in Section \ref{sec:wave}, a sufficiently large phase in the
CKM matrix can only be generated when wave function renormalization
is large which requires $r \sim 4 \pi$. This implies that the one-loop
approximation is not reliable. Therefore, it is most useful to
parameterize the wave function coefficient by an arbitrary
hermitian matrix $Z_{\overline 5}$.

As it stands, this model is incomplete because of the large Yukawa
coupling $r$. The problem is that if the scale of CP violation is
below the Planck scale then the Yukawa coupling runs
to values of order one within one e-folding even if it is
$4\pi$ at the Planck scale. A large
Yukawa coupling at the lower scale $M_{CP}$ can be arranged by
letting the $F_i$'s and $\overline 5_4+ 5_4$ interact with a new
strong gauge group.
It is easy to modify this model to include these interactions.
We present such a model in Appendix \ref{app:nonpert},
where we also show that the relevant non-perturbative effects
in $r$ and the new strong gauge coupling can be determined exactly
and do not contribute to $\tbar$.

\subsection{Reproducing the quark masses and CKM matrix 
\label{subsec:bigckm}}
 
We make some general remarks on models with CP violation from kinetic 
terms.
Since wave function renormalization factors are required to be large
(and not computable in perturbation theory) and flavor violating,
their effects on quark masses and mixing angles are important.
This suggests two different basic scenarios (models which interpolate
between the two extreme cases are of course also possible):

\noindent 
A) The hierarchical structure of the Yukawa couplings is generated
at a scale above $M_{CP}$, and the wave function renormalization
is only responsible for generating the necessary phases. In the
process, the strong dynamics necessarily changes at least some
of the mixing angles completely, but the quark mass hierarchy is
essentially unchanged.

\noindent
B) Flavor and CP violation have a common origin. At scales above
$M_{CP}$ the Yukawa couplings are either universal
($\hat Y^i_j \propto \delta^i_j$) because of non-abelian
flavor symmetries, 
or have ``random'' $O(1)$ entries (flavor anarchy), and the
entire flavor structure including the hierarchy stems from the
wave function renormalization factors $T$. Models in which flavor
originates from wave function renormalization have 
been built by Nelson and Strassler \cite{nelsonstrassler}. Their
models, when adapted to incorporate our mechanism, generate flavor
and solve the strong CP problem.

\subsection{The trouble with models without SUSY}

Note that the necessity of strong coupling $r\sim 4 \pi$
underlines why supersymmetry is so important to our approach:
Non-SUSY models of CP violation induced by non-canonical kinetic terms
have been discussed in the literature, with the new sector coupling 
only to the doublet quarks \cite{GG}, or to the 
singlets \cite{darwin}.
Without SUSY no non-renormalization theorem protects the
colored fermion masses from CP violating vertex corrections, 
which occur at some -- possibly high -- loop level. However, because of
the required large coupling for $r$, arbitrarily high loop
diagrams can still violate the bound on $\tbar$.
Turning the argument around, EDM data put severe constraints on the
model parameters, in particular the coupling $r$.
For example, because of a vertex correction at four loops the authors
of Ref.~\cite{GG} were forced to take $r<1$, and therefore their model
cannot produce large CKM CP violation. Like the one(s)
in \cite{darwin}, it is super-weak and therefore ruled out.
In general, this is the fate of non-SUSY models with
CP violation from kinetic terms; 
supersymmetry and its non-renormalization theorems appear to be necessary
ingredients for this mechanism to yield realistic models with a large
CKM phase. 

\section{SUSY breaking \label{sec:susybreak}}

In this section we discuss the constraints on SUSY breaking and
communication which follow from our solution to the strong CP
problem. 
The non-renormalization theorems guarantee $\tbar=0$ with exact SUSY.
However, after SUSY breaking $\tbar$ is renormalized, and
we find that avoiding large contributions from loops including
superpartners forces the SUSY breaking masses to be highly degenerate.
Furthermore, flavor preserving SUSY parameters, such as the gaugino
masses and $B\mu$ are required to be real to a high accuracy.
We also argue that low-energy SUSY breaking models such as gauge
mediation are most compatible with our mechanism. This is because
the CP violating dynamics renormalizes the SUSY breaking masses
and spoils the necessary degeneracies if soft masses are already present
at the high scale $M_{CP}$.
We give a more detailed discussion of the renormalization of $\tbar$
in gauge mediation in Section \ref{sec:gmsb} and
Appendix \ref{app:radcor} where we also give more references.
 
To begin, note that in the MSSM Eq.~(\ref{eq:thetaeq}) is generalized to
\beq
\tbar=\theta-\arg \det Y_u-\arg \det Y_d-3 \arg (v_u v_d)
       - 3\, \arg\, m_{\tilde g}\ .
\eeq
This immediately implies a strong constraint on SUSY
breaking parameters as the gluino mass and the Higgs vevs have
to be real to one part in $10^{10}$. The reality of the Higgs vevs
translates into constraints on parameters in the Higgs potential.
In particular, a complex $B\mu$ induces complex
vevs already at the tree level. We discuss further constraints
on the Higgs potential in Section \ref{subsec:nonrenorm}.

\begin{figure}[htb]
\vskip 0.0truein
\centerline{\epsfysize=1.5in
{\epsffile{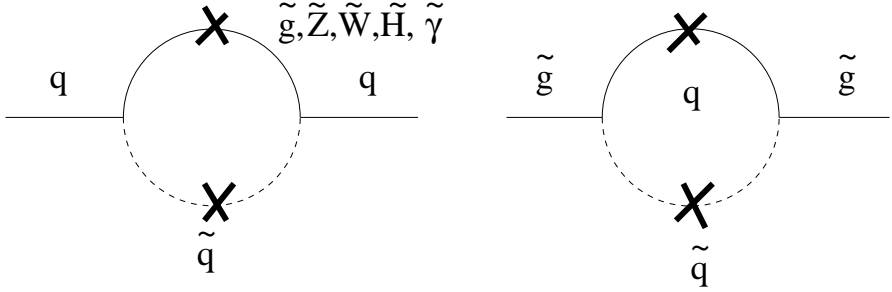}}}
\vskip 0.0truein
\caption[]{\it Lowest order SUSY diagrams contributing to $\tbar$. 
A cross denotes a left-right mass insertion. }
\label{fig:loop}
\end{figure}
Phases of all other flavor-blind MSSM parameters are constrained
because they feed
into colored fermion masses through radiative corrections from 
the diagrams of Figure \ref{fig:loop}.
We summarize these constraints as
\beq
\label{eq:breakbounds}
\arg  m_{\tilde g}, \arg B\mu < 10^{-10} \, , \
\arg  A_0 , \arg  \mu < 10^{-8} \, , \
\arg  m_{\tilde \gamma, \tilde Z,\tilde W }  < 10^{-7}
\eeq
where $A_0$ denotes the proportionality constant of the A-terms
$A=A_0 Y$. We note that these constraints are much more stringent
than the bounds on soft phases from direct contributions
to EDM's which only require phases to be smaller than order $10^{-2}$.

The diagrams in Figure \ref{fig:loop} also lead to strong
constraints on both real and imaginary parts of flavor
violating soft masses. The contributions to $\tbar$ are proportional
to traces over flavor violating quantities such as  
$\Imtr{Y_x^\dagger A_x}$, $\Imtr{Y_x^{-1} A_x}$,
$\Imtr{Y_x^{-1} m^2_{\tilde q} A_x m^2_{\tilde x}}$ where $x=u,d$, see 
Appendix \ref{app:radcor}.
The most natural way to satisfy the bounds on $\tbar$ is to assume 
proportionality and degeneracy
\beq
\label{eq:degenerate}
A_{u,d} \propto Y_{u,d}~,\qquad m_{\tilde{q},\tilde{u},\tilde{d}}^2 \propto 1
\eeq
Deviations from Eq.~(\ref{eq:degenerate}), parameterized 
as $ \delta A$ and $\delta m^2$,
are very constrained, see Appendix \ref{app:radcor}. For generic
deviations which are not ``aligned'' with the Yukawa matrices,
the constraints on some of the matrix elements are as strong as
\begin{eqnarray}
{\delta A \over m_0} < 10^{-13} \hskip1in 
{\delta m^2 \over (m_0)^2} < 10^{-6} \ ,
\label{eq:deltabounds}
\end{eqnarray}
where $m_0$ denotes the average superpartner mass scale. These bounds
apply at $m_0$.
They require a much higher degree of
``flavor blindness'' from the mechanism of
supersymmetry breaking and mediation than FCNC bounds.

Note also that the contributions to $\tbar$ from loops with
flavor violation in superpartner masses
do not decouple in the limit of heavy superpartners. Thus
the bounds (\ref{eq:deltabounds}) apply equally for heavier superpartners.
This is in contrast to the case of FCNCs.

\subsection{Why do we need $M_{CP} > M_{SUSY}$ ?}

For example, minimal supergravity (mSUGRA)
is not compatible with our solution to the strong CP problem;
this can be seen as follows. The Kaehler potential
relevant for squark masses in mSUGRA is
\beq
\int d^4 \theta\  Z_{ij}\, Q_i^\dagger Q_j 
     + {S^\dagger S \over M_{Pl}^2} X_{ij}\, Q_i^\dagger Q_j 
\label{eq:msugra}
\eeq
for the quark $SU(2)$-doublets and similar terms for the singlets.
If we assume a SUSY breaking expectation value for the $F$-component of
the superfield $S$ scalar masses result
\beq
(m_{\tilde q}^2)_{ij}= (T X T)_{ij} {F^* F \over M_{Pl}^2} \ ,
\eeq
where $Z=T^{-2}$ as in Section \ref{sec:wave}. Of course, we can always
work in a basis where $T=1$ at $M_{Pl}$, but in general $X$ will not
be proportional to the unit matrix in this same basis. Partial
alignment $X\simeq Z$ can be achieved by imposing non-abelian
flavor symmetries \cite{nonab}, but residual
non-degeneracies are expected to violate the bounds
Eq.~(\ref{eq:deltabounds}) by orders of magnitude \cite{DKL}.
This is the usual flavor problem of mSUGRA.

In our scenario, the situation for a SUSY breaking mechanism
where superpartner masses are generated at scales above $M_{CP}$
is even worse. This is because the flavor- and CP-violating dynamics at
$M_{CP}$ renormalizes $Z$ and $X$ in Eq.~(\ref{eq:msugra}) differently.
So, even if we had somehow arranged $X=Z$ at $M_{Pl}$, this
alignment would be spoiled at scales below $M_{CP}$.%
\footnote{The only known exception to this is anomaly mediation \cite{AM}
where the special form of supersymmetry breaking proportional
to the conformal anomaly enforces $X=Z$ at all scales.
Therefore, anomaly mediation works very nicely with our scenario,
we briefly discuss it in Section \ref{sec:amsb}.}
To illustrate this point we give the one-loop renormalization of the
right-handed down squark masses in the second model of the previous
Section. Ignoring all coupling constants except $r_{ij}$ this is
$m^2(\mu)/m^2(M_{Pl})\sim (\mu/M_{Pl})^{r^\dagger r /16 \pi^2}$
which is completely non-universal when $r\sim 4\pi$.

We conclude that SUSY models with spontaneous CP violation require
a mechanism of SUSY breaking and mediation in which the scalar masses are
generated below $M_{CP}$. 
We will therefore discuss gauge mediation as a compatible
SUSY breaking mechanism in more detail in the next section.

Of course, any other mechanism of SUSY breaking which generates
universal scalar masses at low scales is compatible with our scheme.
A preliminary look at gaugino mediation \cite{ginom} with
the CPX dynamics at $M_{Pl}$ on the visible sector brane suggests that
gaugino mediation is also compatible with the constraints
Eq.~(\ref{eq:deltabounds}).

\section{Renormalization of $\tbar$ in gauge mediation}
\label{sec:gmsb}

In this section we summarize results on the renormalization of $\tbar$ in
the MSSM with gauge mediated supersymmetry breaking (GMSB).
That gauge mediation
is compatible with solutions to the strong CP problem based on spontaneous
CP violation has been known for some time \cite{gmcp}. We quote here
only the most important results with further details provided in
Appendix \ref{app:radcor}.

In GMSB, the superpartner masses arise from loop
diagrams involving the SM gauge interactions and messenger particles with
SUSY violating masses. The dominant contributions to the scalar masses from
these diagrams have loop momenta of order of the messenger mass;
at higher energies the scalar masses are power-suppressed.
Thus by separating the messenger scale
(which we have been calling $M_{SUSY}$) and CPX scale
$M_{CP}>M_{SUSY}$, one can suppress the dangerous renormalization
of the scalar masses from the CPX sector.

The leading contributions to $\tbar$ in GMSB can be divided into two
classes which we discuss in turn. Contributions which
arise in the effective theory below $M_{CP}$ from renormalizable
interactions and are relatively model-independent and contributions
from higher dimensional operators suppressed by the scale $M_{CP}$
which are model-dependent but can always be made small by
taking $M_{CP} \gg M_{SUSY}$.

At the renormalizable level the only flavor violating couplings
in the effective theory below $M_{CP}$ are the Yukawa couplings.
The gauge-mediated soft SUSY violating masses are approximately given by
$A_x=0$ and $m_{\tilde x}^2=(m_0)^2$. Using the flavor symmetries one can
then show that the renormalization of $\tbar$ from SUSY breaking
can always be written in terms of the hermitian matrices $h_x=Y_x Y_x^\dagger$
in the combination $\det\, \left[h_u,h_d\right]$,
the Jarlskog invariant, see Appendix \ref{app:notation}. 
The leading contribution is
$\delta \tbar \sim 10^{-29}\ tan^6 \beta$ which is smaller than
the leading finite SM renormalization $\delta \tbar \sim 10^{-19}$,
see Appendix \ref{app:radcor}.

The other class of contributions to $\tbar$ involve higher dimensional
operators generated from integrating out the strong CPX
dynamics at $M_{CP}$. For example
\beq
\int d^4\theta {D^\dagger D D^\dagger D \over M_{CP}^2} \ 
{\rm and} \ \left({\alpha_s \over 4 \pi}\right)^2
{X^\dagger X \over M_{CP}^2} D^\dagger D \ .
\label{eq:gmsbkaehlerops}
\eeq
Here $D$ is the right-handed down quark superfield and $X$ is the superfield
whose $F$-component is the source of supersymmetry breaking. 
The first of these operators is generated from the CPX violating dynamics directly
whereas the second comes from computing the two-loop gauge mediation diagram
for scalar masses but restricting the loop momenta to be above the scale $M_{CP}$.
Because of the strongly coupled CPX dynamics at $M_{CP}$ the coefficients of
these operators are not calculable and flavor-off-diagonal. 
Both operators lead to contributions to $\delta m^2$ which are proportional to
$(M_{SUSY}/M_{CP})^2$. The bounds given in
Eq.~(\ref{eq:deltabounds}) then constrain
\beq
\label{eq:softbound}
{M_{SUSY} \over M_{CP}} \lessim 10^{-3}\ .
\eeq
Combining this with the fact that the gauge mediation scale is bounded from
below by roughly $10^4$ GeV we learn that $M_{CP} \gsim 10^7$ GeV, well out
of reach of any current or planned accelerator. 
Both scales are a priori unknown so that we cannot predict the
size of $\tbar$.

We discuss implications
on superpartner masses and electric dipole moments in Section
\ref{sec:predictions}.
For a brief compilation of values of $\tbar$ induced by renormalization
see Table \ref{tab:tbar} in Section \ref{sec:summ}.

\subsection{Anomaly mediation}
\label{sec:amsb}
Even though SUSY breaking and mediation are at high scales
in anomaly mediation (AMSB) \cite{AM},
the superpartner masses at the weak scale are
determined by supersymmetric low-energy
couplings. They are ultra-violet insensitive and therefore
independent of the CPX
dynamics. The resulting soft terms are approximately flavor-universal,
and contributions to $\tbar$ are similar to the contributions from
renormalizable couplings in gauge mediation, negligibly small.
A complete model of course requires a
solution to the problem of negative slepton masses. Any solution which retains the
UV-insensitivity and flavor-universality is compatible with our framework. A
nice example is given by \cite{nimadave}.

\subsection{Contributions from higher dimensional operators}
\label{subsec:nonrenorm}

In this section we discuss a number of different corrections to
$\tbar$ which are model dependent. They include higher dimensional
operators in the superpotential, corrections to the gauge kinetic
functions, Kaehler potential terms which renormalize the superpotential
after SUSY breaking, and higher derivative operators. All of these
operators may arise from quantum gravity dynamics suppressed by the
Planck scale, but some may also arise from CPX dynamics and are therefore
only suppressed by $M_{CP}$.

{\it i. Higher dimensional operators in the superpotential:} 
We showed in Section \ref{sec:susy} that in the absence of SUSY breaking
the only contributions to $\tbar$ can come from the superpotential. Therefore
the most dangerous couplings are direct couplings of CP violating vevs to
the MSSM or any colored fields in the superpotential. In order for our
mechanism to work we must assume that there are no such couplings
at the renormalizable level. For example, we cannot have the couplings
$\Sigma H_u H_d$ or $\Sigma T \overline T$. Both of these couplings can easily
be forbidden by a symmetry under which $\Sigma$ transforms and the MSSM
fields are neutral. At the non-renormalizable level we may have
\beq
\int d^2\theta \ ({\Sigma \over M_{Pl}})^k W_\alpha W^\alpha\ +\ 
       ({\Sigma \over M_{Pl}})^l \mu H_u H_d \ + \
       ({\Sigma \over M_{Pl}})^m Q \hat Y_u U H_u \ .
\eeq

Each of these operators, if present, would give a contribution
to $\tbar$ which is proportional to powers of $M_{CP}/M_{Pl}$ and
could be important if $M_{CP}$ is large and the exponents $k,l,m$ are
small. Again, these superpotential operators are
strongly constrained by symmetries and even in the absence of 
symmetries superpotentials need not be generic
because of the non-renormalization theorems. For example, in our 
model in Section \ref{subsec:ourmodel}, 
one can define a $U(1)$ symmetry under which
only $\Sigma$ and $\overline F$ are charged and which
forbids all these terms.

{\it ii. Kaehler potential terms involving SUSY breaking:} Higher dimensional
operators in the Kaehler potential which couple the MSSM fields to fields with
SUSY breaking vevs can give rise to superpotential terms proportional to 
SUSY breaking. For example, a Kaehler potential term
$X^\dagger/M_{Pl}^2 Q H_d D$ with complex coefficient gives rise to
a superpotential Yukawa coupling with a coefficient $F/M_{Pl}^2$. The same
operator with $M_{Pl}$ replaced by $M_{CP}$ is suppressed by powers
of SM gauge couplings over $16 \pi^2$ if SUSY breaking and CPX dynamics
are not strongly coupled to each other. None of these operators are
dangerous of the SUSY breaking scale is sufficiently low.

{\it iii. Higher derivative Kaehler terms:} Kaehler potential terms
involving the covariant derivative $D_\alpha$ suppressed by $M_{Pl}$ or
$M_{CP}$ generate effective $d^2 \theta$ terms with ordinary derivatives.
For example
\beq 
\int d^4\theta\  {Q H_d (D_\alpha)^2 D\over M_{CP}^2}
       \longrightarrow \int d^2\theta \ Q H_d {\fund\over M_{CP}^2} D\ .
\eeq
These operators can have different flavor structure from the
Yukawa couplings and at one-loop give a flavor non-universal renormalization
of the soft SUSY breaking scalar masses which
are suppressed by $(M_{SUSY}/M_{CP})^2$. We find the same bound as from
the higher dimensional operators in Eq.~(\ref{eq:gmsbkaehlerops}):
$M_{SUSY}/M_{CP}\lessim 10^{-3}$. 

{\it iv. Phases in the SUSY breaking sector:} 
If $\Sigma$ couples directly to the SUSY breaking sector, then
one has to worry about generating a complex SUSY breaking vev $F$. 
A phase in $F$ contributes directly to $\tbar$ via the gluino
mass. It is easy to see that couplings of $\Sigma$ in the superpotential
of the dynamical SUSY breaking sector lead to complex $F$. We therefore
need to forbid such couplings; this can be arranged
in the same way as superpotential couplings of $\Sigma$ to the MSSM
fields can be forbidden.
Phases in the Kaehler potential are less dangerous because of
hermiticity of the Kaehler potential. At tree level, and in looking at
simple toy models we found $F \propto \det Z_{SUSY}$ which is real.
Here, $Z_{SUSY}$ is a wave function renormalization factor in the
Kaehler potential of the SUSY breaking sector.

A more general
analysis of phases in SUSY breaking sectors including loop corrections
is desirable but beyond the scope of this paper.
In any case, such phases can always be avoided by separating the
SUSY breaking and CPX sectors. For example, if
the SUSY breaking sector does not carry the global flavor symmetries
of $\Sigma$, then couplings of $\Sigma$ to the SUSY breaking sector
have to be of the form $Tr(\Sigma^\dagger \Sigma)$ and are therefore real.

{\it v. Phases in the MSSM Higgs sector:}
We already showed that phases in $\mu$ or $B\mu$ are strongly
constrained. $m_{H_u}^2$ and $m_{H_d}^2$ and the supersymmetric quartic
couplings are automatically real, but one might worry about phases
from higher dimensional operators in the Kaehler potential for
$H_u$ and $H_d$. For example,
\beq
\int d^4 \theta {c \over M^2} H_u^2 H_u^\dagger H_d + h.c.
\eeq
with complex $c$ and $M=M_{Pl}$ or $M=M_{CP}$
leads to complex phases in the Higgs vevs which are suppressed
by $(M_{weak}/M)^2$. This is harmless even for the lowest
possible values of $M_{CP}$ and unsuppressed coupling constant $c$.

\section{Predictions}
\label{sec:predictions}

Our framework requires tight constraints on the flavor (and CP)
structure of the SUSY breaking soft terms which
have various testable consequences.
We predict \cite{hilschshort}
\begin{enumerate}
\item Supersymmetry
\item Minimal flavor violation, i.e., there are no significant
new sources of flavor violation beyond the Yukawa couplings at energies
near the weak scale. 
This has well-known implications for $B$-physics \cite{buras2,GGG}.
\item No measurable new CP violation in the quark sector beyond the
SM, in particular no new CP violation in the $B$-system. For example 
$\sin 2 \beta$ is large as in the SM \cite{buras2}. We might expect the
phases in the lepton mixing matrices to be large in 
analogy with the quarks.

\item Almost degenerate first and second generation scalars of each 
gauge quantum number. Generic violations of quark mass universality
are very tightly constrained (see Eq.~(\ref{eq:deltabounds})). However, by
aligning squark masses with quark masses 
\bea
\nonumber
m_{\tilde q}^2&=&m_0^2 \,  (1 + c_u Y_u Y_u^\dagger +c_d Y_d Y_d^\dagger) \, ,\\
m_{\tilde x}^2&=&m_0^2 \, (1 + c_x Y_x^\dagger Y_x)\ ,
\label{eq:addi}
\eea
the renormalization of $\tbar$ remains small ($<10^{-10}$) as can be
seen from Eqs.~(\ref{eq:gluinoA})-(\ref{eq:quarksoft}),
even though this ansatz allows for more flavor violation
than Eq.~(\ref{eq:degenerate}). In this Ansatz the
(real) coefficients $c_{u,d}$ are not expected to be arbitrarily large since
at some point the contribution to 3rd generation superpartners becomes very
large. Imposing $c_i<1/Y_3^2$, where $Y_3$ denotes the
Yukawa of the top, bottom and tau, gives $\triangle m < 1$ GeV for
the difference between the first and second generation scalars.
This is a prediction which should be tested at a linear collider. 
We stress that this degeneracy holds independent of the SUSY
breaking mechanism. It follows only from demanding that the
radiative corrections to $\tbar$ not be too large (and a reasonable
constraint on the $c_i$).
Note that this also bypasses possible FCNC problems since
the resulting off-diagonal squark masses obey
\beq
\frac{\triangle m_{\tilde q}^2}{m_0^2} \phantom{|}_ {(12,13,23)}
     \leq (V_{ub} V_{cb}^*,V_{ub} V_{tb}^*,V_{cb} V_{tb}^*)
     \lessim (10^{-5},10^{-2},10^{-1})
\eeq
where $\triangle m_{\tilde q}^2\phantom{|}_{(ij)}$ denotes the mixing 
between the
$i$th and $j$th generation.
These values are within the experimental bounds \cite{susyfcnc}.

\item At the renormalizable level, 
the radiatively induced strong CP phase is of the order 
$\tbar \simeq 10^{-19}$.
However, depending on the model dependent ratios
$M_{SUSY}/M_{CP}$ and
$M_{CP}/M_{Pl}$, the strong phase $\tbar$ can be as large as $10^{-10}$.
Thus,
the corresponding hadron electric 
dipole moments can be close to the experimental bound and
might be measured soon \cite{golublamoreaux}. 

\item
A weak electric dipole moment
$d_f$ is a contribution to the 5-dimensional operator
$d_f \frac{i}{2}\bar{f} \sigma_{\mu \nu} \gamma_5 f F^{\mu \nu}$.
The EDMs for quarks and leptons arise from 1-loop diagrams
like Figure \ref{fig:loop} with
an external photon attached to wherever possible.
In the general MSSM with arbitrary phases the experimental bound
from the electron EDM
$d_{e} < 1.8 \cdot 10^{-27} e cm$ \cite{PDG}, and from the neutron EDM
$d_{n} < 6.3 \cdot 10^{-26} e cm$ \cite{PDG} require the phases of
$\mu$, $A$-terms and the gaugino masses to be less than $10^{-2}$, 
e.g.~\cite{DGH}.
Since the dipole moments are linear in the soft SUSY phases we
conclude that the phases which are constrained by Eqs.~(\ref{eq:breakbounds})
and (\ref{eq:deltabounds}) give weak quark and lepton EDMs
which are at least
five orders of magnitude below their experimental bounds.
Note that improvements of the experimental EDM limits further
strengthen the bounds Eq.~(\ref{eq:breakbounds}),
thus weak EDMs are always smaller than strong EDMs in our framework.

\item
Large flavor-preserving phases in the soft terms 
with their associated ``SUSY CP-problem'' have no place in our
framework, see the bounds in Eq.~(\ref{eq:breakbounds}).
This simplifies in particular the 
phenomenological analysis of the Higgs potential.

\end{enumerate}

\section{Summary and concluding remarks \label{sec:summ}}

We presented a new theory of CP with supersymmetry and spontaneous
CP violation. CP is assumed to break spontaneously and CP violation
is communicated to the MSSM fields at the scale $M_{CP}$. 
SUSY breaking is communicated to the MSSM at the lower scale $M_{SUSY}$.
With these ingredients, a natural solution to the strong CP problem arises, 
because at the scale of CP violation the strong CP phase $\tbar$ is protected
by a non-renormalization theorem of the unbroken supersymmetry. At lower
energies SUSY is broken and the non-renormalization theorem does not
apply, but we showed that the generated $\tbar$ is much smaller
than the experimental bound if SUSY breaking is sufficiently flavor-universal.
Because of the non-renormalization theorem at high scales a
successful model for the CP violating sector only needs to ensure
$\tbar=0$ at the tree level which is easy to arrange. 
The CKM phase is generated either at the tree level as in 
Nelson-Barr models or else at the loop level from wave function
renormalization as we proposed in \cite{hilschshort}.

We have explicitly shown that low scale gauge mediation with 
$M_{SUSY} < M_{CP}$
is compatible with our framework, but other mechanisms can also be implemented.
A model independent constraint is that SUSY breaking has to be CP conserving
and either flavor-universal or else flavor-aligned as in Eq.~(\ref{eq:addi}).
A summary of values of $\tbar$ in some theories discussed in this paper
is compiled in Table \ref{tab:tbar}.

\begin{table}[ht]
\renewcommand{\arraystretch}{1.8}
         \begin{center}
         \begin{tabular}{|c|c|c|c|c|}
  \hline
     \multicolumn{1}{|c|}{\mbox{}}
       & \multicolumn{1}{|c|}{$SM$}
       & \multicolumn{1}{|c|}{$MSSM_{gen}$}
       & \multicolumn{1}{|c|}{$MSSM_{flav}$}
       & \multicolumn{1}{|c|}{$HDO_{GMSB}$} \\
         \hline
$\tbar$ & $\sim 2 \cdot 10^{-19}$ &
$\frac{\displaystyle\alpha_s}{\displaystyle 4 \pi} $ &
 $\sim 2 \cdot 10^{-19}$ & $
(\frac{\displaystyle \alpha_s}{\displaystyle 4 \pi})^2
         (\frac{\displaystyle M_{SUSY}}{\displaystyle M_{cp}} )^2$ \\
\hline
$\delta \tbar_{RGE}$ &
$ 10^{-30}$ & & $10^{-29} \tan(\beta)^6$ &  \\
         \hline
         \end{tabular}
         \end{center}
\caption{\it Magnitude of $\tbar$ from renormalization starting from
$\tbar_{tree}=0$ in some theories discussed in text. Here, SM denotes the
Standard Model (SM)
and $MSSM_{gen}$ a generic minimal supersymmetric model.
In the minimal supersymmetric model denoted as $MSSM_{flav}$
flavor violation is minimal, i.e., not bigger than
in the SM. This suppresses large radiative corrections
to $\tbar$ that are present in $MSSM_{gen}$.
Note that the MSSM with gauge mediated SUSY breaking (GMSB)
belongs to class $MSSM_{flav}$.
The last column corresponds to contributions from
higher dimensional operators (HDO) in GMSB. Now
the size of $\tbar$ depends on the hierarchy between the scale of SUSY
breaking $M_{SUSY}$ and the
scale of spontaneous CP violation $M_{CP}$.
The last line shows the contributions to $\tbar$ from RGE running
in the SM and $MSSM_{flav}$.}
\label{tab:tbar}
\end{table}

{}From the low energy point of view, our theory is the MSSM
with miniscule flavor violation and no significant phases beyond
those already present in the SM.
The only possible deviation from this picture 
is that higher dimensional operators may
bring the nucleon EDMs into experimental reach.
Our proposal requires supersymmetry, and the strong constraints on
the superpartner spectrum from the renormalization of $\tbar$
automatically also nullify the SUSY phases and FCNC
problems. We have pointed out many of the
testable signatures for $B$-physics, collider and nucleon 
EDM experiments.
Note that our proposal does not require light superpartners; by
low scale SUSY breaking we mean that its {\it mediation} to the MSSM occurs
below $M_{CP}$.

In our paper we have given two explicit examples of CP violence, but
we stress that our solution to the strong CP problem can be
incorporated in a much larger
class of models because our main tool, 
the non-renormalization of $\tbar$ in SUSY, is general.
It would be interesting to combine our theory of CP with a theory
of flavor, e.g.~with \cite{nelsonstrassler}. 
This is because a necessary ingredient in our CP violating sectors
is flavor violation. 
Thus there may be elegant models in which
both goals are achieved at one. Such a model could also include
grand unification.

Finally, we briefly comment on cosmological issues.
The spontaneous breaking of CP
leads to the formation of domain walls. Such domain walls are potentially
problematic because they can over-close the universe. However, in our
theory the scale of CP-breaking is sufficiently high that several possible
mechanisms (including inflation) exist to avoid this problem.
Baryogenesis can occur in a number of different ways such as
CP-asymmetrical decays of GUT-scale or $M_{CP}$-scale particles,
the Affleck-Dine mechanism, or Leptogenesis. 
 
\vskip.4in
{\bf \large \noindent Acknowledgements}
\vskip.2in
 
\noindent We thank Bill Bardeen, Wilfried Buchm\"uller, 
Koichi Hamaguchi, David E.~Kaplan, Ann Nelson,
Michael Peskin, and Tsutomu Yanagida
for stimulating questions and discussions. 
MS was supported by the
Department of Energy under contract DE-AC02-76CH03000.

%
%
\begin{appendix}
\renewcommand{\theequation}{\Alph{section}-\arabic{equation}}

\setcounter{equation}{0}
\section{Notation \label{app:notation}} 

We settle here our notation of quark masses and the CKM mixing matrix
$V_{CKM}$. 
\begin{eqnarray} 
M_u=diag(m_u,m_c,m_t) ~,&&\qquad 
M_d=diag(m_d,m_s,m_b) \\
M_u=V_u Y_u U_u^{\dagger} v_u ~,&&\qquad 
M_d=V_d Y_d U_d^{\dagger} v_d 
\end{eqnarray} 
We will also use the normalized mass matrices
\beq
\hat M_u=diag(\frac{m_u}{m_t},\frac{m_c}{m_t},1) ~~~~,~~
\hat M_d=diag(\frac{m_d}{m_b},\frac{m_s}{m_b},1) \; .
\eeq
Here, the unitary matrices, $U_{u,d},V_{u,d}$ diagonalize the
Yukawas $Y_{u,d}$, which are
given in the basis with canonical kinetic terms.
\begin{eqnarray}
\label{eq:ckm1} 
Y_u Y_u^{ \dagger} v_u^2& =&V_u^{\dagger} M_u^2 V_u \\ 
Y_d Y_d^{ \dagger} v_d^2& =&V_d^{\dagger} M_d^2 V_d \\ 
V_{CKM}&=&V_u V^{\dagger}_d 
\label{eq:ckmdef}
\end{eqnarray} 
The amount of weak CP violation in the SM is given by the Jarlskog 
determinant
\begin{eqnarray} 
\det[h_u,h_d]_{SM}&=&2 i J (m_t^2-m_c^2) (m_t^2-m_u^2) 
(m_c^2-m_u^2) \nonumber \\&&(m_b^2-m_s^2)(m_b^2-m_d^2)(m_s^2-m_d^2)/v^{12} 
\end{eqnarray} 
Here, $v=174$ GeV, $h_u=Y_u Y_u^{\dagger}$, $h_d=Y_d Y_d^{ \dagger}$,
$J=s_{12}s_{13} s_{23} c_{12} c_{13}^2 c_{23} \sin \phi_{CKM} $
and $s_{ij}=\sin \phi_{ij}$, $c_{ij}=\cos \phi_{ij}$ where
$\phi_{ij}$ and $\phi_{CKM}$ are the angles and phase of the CKM matrix 
in PDG parameterization. Numerically, $J \approx 2 \cdot 10^{-5}$ \cite{PDG}.

\setcounter{equation}{0}
\section{The CKM phase \label{app:ckm}} 

In this Appendix, we show that the heavy sector has to couple strongly
to the SM fermions to yield an ${\cal{O}}(1)$ CKM phase
from CP violation in the quark kinetic terms.
In particular, the ansatz $Z^{-1/2}=1+\epsilon H$, where
$H$ is hermitian and has order 1 entries, leads to the
observed pattern of quark masses, mixing and CP violation only if
the parameter $\epsilon \gsim 1$.

To begin, we note that if the initial Yukawas do not have the
right (hierarchical) eigenvalues, then large rescaling is required from
the wave function renormalization, which implies $\epsilon \gsim 1$
(We give a proof for this further down below).
Thus, we only have to exclude the possibility that the Yukawas $\hat Y$
already have approximately the correct eigenvalues to correspond
to the SM quark masses but that the CP phase (and possibly also the mixing
angles) are generated from wave function renormalization  with small
$\epsilon$. 
Without loss of generality, we work in a basis in which
$\hat Y_u \approx M_u / v_u$ is diagonal.
It is furthermore general to choose $\hat Y_d \approx O M_d/v_d$
where $O$ is a general orthogonal (real) matrix. Finally, since we
are only concerned with determining the CKM matrix, we are
free to re-scale $\hat Y_u$ and $\hat Y_d$ such that the largest eigenvalue
in each is approximately equal to one.

The CKM matrix is then the unitary transformation between the basis
in which the following two matrices $h_u, h_d$ are diagonal
\bea
h_u={1\over \sqrt{Z_Q}} \, \hat M_u
\frac{\displaystyle 1}{ \displaystyle  Z_u}  \hat M_u \,
{1\over \sqrt{Z_Q}} ~~, ~~
h_d={1\over \sqrt{Z_Q}} \, O \, 
\hat M_d
\frac{\displaystyle 1}{ \displaystyle  Z_d} 
\hat M_d \, O^T \,
{1\over \sqrt{Z_Q}} 
\eea
Now we assume that $\epsilon$ is small and show that one cannot generate
a sufficient amount of CP violation.
First, note that $V_{CKM}=O$ if all $Z_i=1$. Anticipating this to still
be approximately true when the $Z_i$ differ from 1 perturbatively, we
rotate $h_d$ by $O$ so that it's unperturbed component
is already diagonal. We now have 
\beq 
V_{CKM}=V_u O (V_d^O)^\dagger
\eeq
where $V_u$ diagonalizes $h_u$ and $V_d^O$ diagonalizes
\beq
h_d^O=(O^T {1\over \sqrt{Z_Q}} \, O) \, 
\hat M_d
\frac{\displaystyle 1}{ \displaystyle  Z_d} 
\hat M_d \, (O^T \,
{1\over \sqrt{Z_Q}}O)\ .
\eeq

In order to determine the eigenvalues and unitary matrices $V_u$ and $V_d^O$,
we use standard non-degenerate perturbation theory familiar from
quantum mechanics. First, we parameterize
$(Z_Q )^{-1/2}=1+\epsilon H$ and $Z_{u,d}^{-1}=1+\epsilon J_{u,d}$. To linear
order in $\epsilon$ we have
\beq
h_u=\hat{M}_u^2+\epsilon \triangle_u \ , \qquad h_d^O=\hat{M}_d^2+\epsilon \triangle^O_d\ ,
\eeq
where 
\bea
\label{eq:tri}
\triangle_u = \{ H , \hat M_u^2 \}  + \hat M_u J_{u}\hat M_u ~,~~~
\triangle_d^O =\{ H^O , \hat M_d^2  \} +  \hat M_d J_{d} \hat M_d\ ,
\eea
and $H^O=O^T H O$. Here, the unperturbed ``Hamiltonians'' $\hat{M}_u^2$ and $\hat{M}_d^2$
are already diagonal.
The perturbed eigenvalues to order $\epsilon$ are then
\bea
\label{eq:shifts}
(\hat{M}_u^2)_i +\epsilon (\triangle_u)_{ii}
          =  (\hat{M}_u^2)_i \left[1+ \epsilon(2 (H_u^O)_{ii}+(J_u)_{ii})\right]\ ,
\eea
and a similar expression for the down sector. Thus we see that the renormalizations of
individual quark masses are multiplicative, this implies e.g. that there are
no corrections to $m_u$ proportional to $m_t$. Here, we discovered this
property to linear order in $\epsilon$, it is straightforward to
extend this analysis to higher order. We have computed the corrections
up to second order and also verified our results numerically without
expanding in $\epsilon$.
This verifies our claim that large corrections to masses can only come
from non-perturbatively large $\epsilon$.  

The unitary matrices which diagonalize $h_u$ and $h_d^O$ are
\bea
\label{eq:vup}
(V_u)_{ij}=\delta_{ij}
      +\epsilon \, \left.{(\triangle_u)_{ij}\over (\hat M_u^2)_i-(\hat M_u^2)_j}\right|_{i\neq j} \\
\label{eq:vdown}
(V_d^O)_{ij}=\delta_{ij}
      +\epsilon\, \left. {(\triangle_d^O)_{ij}\over (\hat M_d^2)_i-(\hat M_d^2)_j}\right|_{i\neq j} 
\eea

Since contributions to the CKM angles from the different terms above
are additive in perturbation theory
(i.e. $\Pi_i(1+\epsilon_i)=1+ \Sigma_i\epsilon_i$), we discuss each
of them in turn. 

Non-vanishing $J_d$ (contributions from $J_u$ are smaller) in Eq.~(\ref{eq:vdown})
lead to complex corrections to the CKM matrix elements of order
\beq
\label{eq:small}
\delta V_{ub} \sim \epsilon\, \frac{m_d}{m_b} \ , \quad
\delta V_{cb} \sim \epsilon\, \frac{m_s}{m_b} \ , \quad
\delta V_{us} \sim \epsilon\, \frac{m_d}{m_s} \ .
\eeq
This is most significant for $\delta V_{cb}$ and gives
$\phi_{CKM}\lessim \epsilon \frac{m_s}{m_b}/V_{cb} \sim \epsilon$.

The case of non-trivial $Z_Q$ (i.e. non-vanishing H) is slightly more
complicated. Assuming that the matrix $H$ has entries of order one,
and choosing the angles in $O$ similar to the experimental values in $V_{CKM}$
we find for the Jarlskog invariant (see Appendix \ref{app:notation})
\beq
J \lessim 2 \epsilon (\theta_{12} \theta_{23} - \theta_{13}) 
\theta_{13} \theta_{12}\ ,
\eeq
where $\theta_{ik}$ are the angles of $O$ in the parameterization of
the PDG \cite{PDG}. We extract $\sin \phi_{CKM}$
by dividing by the angles. This yields the bound 
\beq
\sin \phi_{CKM} \lessim 2 \epsilon |V_{ub}|/|V_{cb}|
\eeq
which is too small since data imply $\sin \phi_{CKM}\sim {\cal{O}}(1)$.
%

A comment on the usefulness of our expansion in $\epsilon$ is in order:
There are many small parameters in the problem with the
potential danger of factors such as $m_t/m_u$ ruining the
expansion. We believe that such factors do not occur. This
is manifest to order $\epsilon$ from our expressions above, and
we have verified it explicitly to second order. 
Furthermore, extensive numerical study \cite{hillerschmaltzware}
has shown that our results are not affected by higher order
corrections in $\epsilon$:
large departures from canonical kinetic terms are
required if we want to generate sufficient CKM CP-violation
from wave function renormalization. 

\setcounter{equation}{0}
\section{Strong interactions at $M_{CP}$  \label{app:nonpert}} 

The model of Section \ref{subsec:ourmodel}
is incomplete because renormalization
from the Planck scale to $M_{CP}$ drives the Yukawa coupling $r$ 
to values which are too small to give sufficient CP violation
in the quark kinetic terms. The model can be fixed
by introducing a new gauge group $SU(N)$ under which $\overline{5_4}$ and
$F$ transform in the fundamental representation and $5_4$
and $\overline{F}$ are anti-fundamentals. The superpotential
(\ref{eq:superpot}) remains invariant. The $SU(N)$ theory has
8 flavors (5 from $\overline{5_4}+5_4$ and 3 from $F+\overline{F}$)
and its gauge coupling becomes strong in the IR for
$N\ge 3$. The strong gauge interactions then also drive the
Yukawa coupling $r$ to large values as can be seen from
the sign of the beta function (schematically, ignoring coefficients)
\beq
\label{betaforr}
16 \pi^2\ {d \over d (log \mu)}\ r =  r \ \left( r^2 - g_N^2\right)
\eeq
where $g_N$ is the coupling of the new strong $SU(N)$.

At the scale $M_{CP}$, the $F$'s and $5_4$'s are massive.
Integrating them out leads to non-canonical CP violating kinetic
terms for the right handed down quarks (and lepton doublets),
vanishing $\tbar$ and no new superpotential couplings to all orders in
perturbation theory as described in Section \ref{subsec:nonrenorm}.

But what about non-perturbative effects which could
arise from the strong $SU(N)$ dynamics? These effects can
be deduced from Seiberg's solution of supersymmetric
QCD \cite{IS}, \cite{Graesser:1998df}. Most important here are the
matching relations for the strong interaction scale across
mass thresholds. After integrating
out $F$'s and $5_4$'s the $SU(N)$ gauge theory is flavor-less and
confines. Gaugino condensation generates a superpotential
$W = \Lambda_{IR}^3 = ( \Lambda_{UV}^{3N-8} M^5 {\rm det}(\Sigma))^{1/N}$
where $M$ and $\Sigma$ are defined in Section \ref{subsec:ourmodel}, 
and $\Lambda_{UV/IR}$
is the ``QCD'' scale of the one-loop $SU(N)$ beta function
below/above $M_{CP}$.
This superpotential is complex, but it does not couple to any MSSM fields and
is therefore harmless. We should also worry about direct
non-perturbative contributions to $\theta$ of the GUT $SU(5)$ group.
These can be determined from the $SU(5)$ scale matching.
The phase of the scale of the $SU(5)$ group at the high scale
$\Lambda_{5\,UV}^{8-N}$ vanishes because of CP invariance.
This is the statement that $\theta=0$ at the Planck scale.
At lower scales, after integrating out $F$'s
and $5_4$'s, the phase is determined by scale matching:
$\Lambda_{5\,IR}^{8}=\Lambda_{5\,UV}^{8-N} M^N$.
This is also real. At even lower scales the dynamics of the $SU(N)$
theory and the $SU(5)$ are completely decoupled so that no further
scale matching for the $SU(5)$ theory is required. This proves that
$\tbar=0$ in the effective supersymmetric theory below $M_{CP}$
even after including non-perturbative dynamics in the strongly
coupled $SU(N)$ and the coupling $r$.

\setcounter{equation}{0}
\section{Radiatively generated strong CP phase \label{app:radcor}} 

We start with a discussion of contributions to $\tbar$
from the renormalization of
quark masses in the SM.
Corrections can be written as $m=m_0(1+x)$ and we will use
$\arg \, \det(1+x)=\Imtr x$ for small $x$. 
Using the flavor symmetries, it is easy to show that corrections
to $\tbar$ can always be written as the imaginary part of traces over the
hermitian matrices 
$h_u=Y_u Y_u^{\dagger}$ and
$h_d=Y_d Y_d^{\dagger}$
(here we work in the basis with canonical kinetic terms).
The lowest order non-vanishing contribution to $\tbar$ arises at
6th order in $h_{u,d}$. It is related to the Jarlskog determinant
(see Appendix \ref{app:notation})
by 
\beq
\label{eq:imtrdet}
2\, \Imtr{h_u h_d h_u^2 h_d^2}=\det[h_u,h_d]
\eeq
Expressions involving $n$ powers of $h$ arise from diagrams with
at least $n$ loops. Alternatively, they arise in a step-wise linear
approximation to the RGEs
with at least
$n$ steps \cite{DGH}.
This defines our power counting:
1 Higgs loop or 1 RGE step both give $h_{u,d}/(16 \pi^2)$.
Higher orders in $n$ are suppressed and one can show that
Eq.~(\ref{eq:imtrdet}) is indeed the trace with the largest
imaginary part.
But at 6-loops a cancellation occurs between diagrams where up and down
quarks are interchanged because
$\Imtr{h_u h_d h_u^2 h_d^2}+ (u \leftrightarrow d)=0$. An extra 
loop with a photon splits the isospin symmetry.
Thus, the RGE induced
correction to $\tbar$ in the SM is \cite{EG,DGH}
\begin{eqnarray} 
\label{eq:SM}
\theta_{SM}^{RGE} \approx \frac{\alpha}{4 \pi}  
\left(\frac{\triangle t}{16 \pi^2} \right)^6 
\det[h_u,h_d]_{SM} 
\end{eqnarray} 
which is approximately $\theta_{SM}^{RGE} \approx 10^{-30}$
for $\triangle t=log\, M_{Pl}/M_Z$.

The largest contribution to $\tbar$ in the SM arises from the finite
and strongly GIM-suppressed four-loop cheburashka diagram \cite{khriplovich} 
\begin{eqnarray} 
\label{eq:thetasmfinite}
\theta_{SM}^{finite}  
=-\frac{7}{9} \frac{\alpha_s}{4 \pi} \left( \frac{\alpha_W}{4 \pi} \right)^2  
\frac{m_s^2 m_c^2}{m_W^4} J \ln\frac{m_t^2}{m_b^2}
\ln^2 \frac{m_b^2}{m_c^2} \left( \ln\frac{m_c^2}{m_s^2}+\frac{2}{3}
\ln\frac{m_b^2}{m_c^2} \right) 
\end{eqnarray} 
which gives $\theta_{SM}^{finite} \approx 2 \cdot 10^{-19}$ using
$\alpha_s=0.2$ and $J= 2 \cdot 10^{-5}$ and is
consistent with earlier estimates~\cite{EG}. 

In the MSSM, the leading divergent diagrams which renormalize
$\tbar$ cancel because of the SUSY non-renormalization theorem.
However, there are new
finite contributions from one-loop quark and gluino mass corrections
which involve supersymmetry breaking. 
The diagrams for gluino and quark mass renormalization are proportional
to soft $A_x$ terms and soft masses
\footnote{We use the soft Lagrangian as $-{\cal{L}}_{soft} \supset
\tilde{Q} A_u H_u \tilde{U} +\tilde{Q} A_d H_d \tilde{D} +
1/2 m_{\tilde{g}} \tilde{g} \tilde{g} + B\mu H_u H_d+ c.c. +
\tilde{Q}^\dagger m_{\tilde{q}}^2\tilde{Q}+
\tilde{U}^\dagger m_{\tilde{u}}^2\tilde{U}+
\tilde{D}^\dagger m_{\tilde{d}}^2\tilde{D}$, see e.g.~\cite{rgemssm}.}  
$m_{\tilde x}^2, m_{\tilde{q}}^2$ and yield ($x=u,d$)
\bea
\label{eq:gluinoA}
\theta^A_{\tilde{g}} & \simeq & \frac{\alpha_s}{4 \pi} 
\frac{v_x^2}{m_0^3}\,
\Imtr{ Y_x A_x^\dagger } \\
\theta^{m}_{\tilde{g}} & \simeq & \frac{\alpha_s}{4 \pi} 
\frac{v_x^3 v_y}{m_0^{8}}\,
\Imtr{ h_x Y_x m_{\tilde x}^2 Y_x^\dagger m_{\tilde q}^2 } 
\eea
and
similar expressions for quark mass contributions
\bea
\label{eq:quarkA}
\theta^A_{\tilde{q}} & \simeq & \frac{\alpha_s}{4 \pi} 
\frac{1}{m_0}\,
\Imtr{ Y_x^{-1} A_x } \\
\theta^{m}_{\tilde{q}} & \simeq & \frac{\alpha_s}{4 \pi} 
\frac{v_y}{ m_0^{4} v_x}\,
\Imtr{ Y_x^{-1} m_{\tilde q}^2 Y_x m_{\tilde x}^2} 
\label{eq:quarksoft}
\eea
Here $m_0$ is an effective average soft mass, $v_x$ are
the Higgs vevs and $y \neq x$. 
The size of the induced $\tbar$ depends crucially on the 
flavor structure of the soft breaking parameters.
Arbitrary A-terms and soft masses can violate the 
experimental bound on $\tbar$ 
by many orders of magnitude.
On the other hand, for soft terms which satisfy exact
proportionality and degeneracy as in Eq.~(\ref{eq:degenerate})
these contributions to $\tbar$ vanish. However, proportionality and
degeneracy are not stable under renormalization. The RGEs
for the soft terms \cite{rgemssm} involve products of $h_u$ and $h_d$.
Inserting the renormalized soft masses into the one-loop
diagrams Figure \ref{fig:loop}, and using
arguments very similar to the SM discussion above one finds \cite{DGH} 
\begin{eqnarray} 
\label{eq:MSSM}
\theta_{SUSY}^{RGE} \approx \frac{\alpha_s}{4 \pi} 
\left(\frac{\triangle t}{16 \pi^2} \right)^5
\frac{v_u^2 }{m^2_0}
tan^6 \beta \, \det[h_u,h_d]_{SM} 
\end{eqnarray} 
This gives  
$\theta_{SUSY}^{RGE} \approx 10^{-29} - 10^{-19}$ for $tan \beta$
ranging from 1 to 50.

Thus in the MSSM with strictly proportional and universal soft terms at
a high scale (e.g.~$M_{SUSY}$ in GMSB) the contributions from diagrams
involving superpartners are smaller than the finite diagram in the SM.
Diagrams which are similar to the leading SM contribution
Eq.~(\ref{eq:thetasmfinite}) but involve superpartners or charged Higgses
are suppressed by the heavier superpartner and Higgs
masses and are therefore smaller than Eq.~(\ref{eq:thetasmfinite}).

Let us work out the constraints on A-terms and soft masses if we allow
for additional flavor violating contributions. We parameterize
the departure from proportionality and degeneracy as
$\delta A,\delta m^2$.
{}From Eq.~(\ref{eq:quarkA}) follows  immediately for the A-terms
\beq
\Imtr{ Y^{-1} \frac{\delta A}{m_0}} \leq 10^{-8}
\eeq
We need non-universality for both soft masses in Eq.~(\ref{eq:quarksoft})
for a non-zero contribution to $\tbar$.
For example, our power counting discussed previously
gives $m_{\tilde q}^2 \simeq m_0^2 (1 + h_x/(16 \pi^2))$
thus
\beq
\Imtr{ Y_y^{-1} h_x Y_y  \frac{\delta m^2}{m_0^2}} \leq 10^{-6}
\eeq
These bounds are generally much more severe than the bounds
from FCNCs, see e.g \cite{susyfcnc}.
The constraints on the smallest elements of  $\delta A$ and $\delta m^2$
are quoted in Eq.~(\ref{eq:deltabounds}).

\end{appendix}

\end{document}